# Engineering of Microfabricated Ion Traps and Integration of Advanced On-Chip Features

*Zak David Romaszko*, Seokjun Hong, Martin Siegele, Reuben Kahan Puddy, Foni Raphaël Lebrun-Gallagher, Sebastian Weidt, Winfried Karl Hensinger*

*Sussex Centre for Quantum Technologies, Department of Physics and Astronomy, University of Sussex, Brighton, BN1 9QH, United Kingdom*

## Abstract

Atomic ions trapped in electromagnetic potentials have long been used for fundamental studies in quantum physics. Over the past two decades trapped ions have been successfully used to implement technologies such as quantum computing, quantum simulation, atomic clocks, mass spectrometers and quantum sensors. Advanced fabrication techniques, taken from other established or emerging disciplines, are used to create new, reliable ion trap devices aimed at large-scale integration and compatibility with commercial fabrication. This Technical Review covers the fundamentals of ion trapping before discussing the design of ion traps for the aforementioned applications. We overview the current microfabrication techniques and the various considerations behind the choice of materials and processes. Finally, we discuss current efforts to include advanced, on-chip features into next generation ion traps.

## 1. Introduction

The trapping of atomic ions in confining electric fields in vacuum was first conceived and demonstrated by Wolfgang Paul and Hans Georg Dehmelt, earning them a share of the 1989 Nobel prize in physics [1], [2]. An ion isolated in this way can be extremely well decoupled from its environment and thus cooled to very low temperatures using laser techniques (such as those in Ref. [3]). The extreme isolation and low thermal energy mean that the energy levels of the laser-cooled ion are highly stable and well resolved, with quantum states having been observed to remain coherent over several minutes [4], [5]. These properties, along with the ability to prepare and detect the quantum states and generate high fidelity entanglement between trapped ions, make trapped ion systems particularly well-suited for experiments that require the precise control of well-defined quantum systems such as atomic clocks [6], quantum sensors [7], quantum simulators [8]–[11], mass spectrometers [12]–[14] and quantum computers [15], [16].

The traditional ion trap design introduced by Paul, the Paul trap, uses oscillating (RF) voltages to create potential minima in up to three dimensions (Fig. 2.1(a)), which, when combined with static DC fields, are able to confine the ion to a certain position in space [1]. With numerous applications of trapped ions, comes the desire to significantly increase the number of ions while maintaining, and in some cases increasing, the precise control over the position of individual ions. Quantum computing is a good example where many approaches to scalability require such a level of control [17]–[19]. The precise control of the ion positions requires a significant reduction in the size and an increase in the number of control electrodes, making the early type of Paul traps, consisting of mechanically machined 3D electrode structures, unsuitable [20]. The use of microfabrication methods allows for the realisation of the required feature sizes, reproducibility and mass producibility needed for such devices. This led to an early prototype, in 1999, which was made using laser-machined, gold-on-alumina wafers [21]. It was followed by monolithic ion microchips which were lithographically patterned, first demonstrated in 2006, Refs. [22], [23]. Whilst quantum computing provided the first motivation for using microfabricated devices in ion traps, other trapped-ion technologies took advantage of these developments. For instance, lithographic methods contributed to the realisation of a compact ion trap package for portable atomic clocks [24], [25] and also quantum simulations in 2D lattices with micron-scale inter-ion separation [26], [27].

*Z.Romaszko@sussex.ac.uk

These 'trap-on-chip' devices use established fabrication techniques from the semiconductor and micro-electro-mechanical system (MEMS) industries to realise micron-scale architectures in both 2D (surface traps) [23], [28] and 3D configurations [22], [29]. Another set of well-established techniques, which is also highly prevalent in modern electronics, comes from complementary metal–oxide–semiconductor (CMOS) technology and has been used to successfully fabricate an ion trap [30]. These methods have allowed the creation of reliable ion traps using established processes, giving trapped ions a prominent position in quantum technologies.

In addition to the miniaturisation of ion traps, the integration of peripheral components, such as photodetectors [31] and digital to analogue converters (DACs) [32], into the ion trap is also of great interest because it allows the creation of compact devices and stand-alone trap modules and has also the potential to reduce electrical noise [33]. The integration of peripheral components is also critical to large scale quantum computing with trapped ions where stand-alone modules are a key ingredient to scalability [18], [19].

To create a quantum computer with sufficient number of qubits to perform interesting operations, the ability to connect different individual modules is required. Two methods that address this connectivity have been proposed. One scheme [18], [34], uses photons, emitted by ions in separate modules, to create inter-module entanglement. Another method [19], relies on shaping electrodes in such a way that when neighbouring modules are closely aligned, ions can be transported from one module to another using electric fields.

## BOX 1: Ion trapping basics

The application of an RF voltage to the ion trap electrodes, creates a trapping potential. In addition, DC voltages produce a static field, that confines the ion in the axial direction [1]. This can be expressed as,

$$\varphi_{tot}(x,y,z,t) = \varphi_{DC}(x,y,z) + \varphi_{RF}(x,y,z)\cos(\Omega t)\,, \quad (1)$$

where $\varphi_{tot}$, $\varphi_{DC}$ and $\varphi_{RF}$ is the total, DC and RF potentials respectively and Ω is the frequency at which RF is being driven.
The equations of motion for a particle of mass *m* in a Paul trap are given by,

$$\ddot{x} + \frac{e}{mr_0^2}(\varphi_{DC} - \varphi_{RF}\cos(\Omega t))x = 0\,. \quad (2)$$

where $r_0$ is the distance from the centre of the trap to the RF electrodes. These equations can be written in the form of the Mathieu equations

$$\frac{d^2 i}{d\zeta^2} + (a_i - 2q_i\cos(2\zeta))i = 0\,, \quad (3)$$

where $a_i$ and $q_i$ are the stability parameters in a direction, *i*. Only a small subset of these parameters provide stable trapping which are given by the Floquet theorem [86]. In the regime where $1 > q_i >> a_i$, the ions motion can be characterised in two ways, 'secular motion' and 'micromotion'. The secular motion is the motion due to the curvature of the electric potential. This motion is often used to implement spin-motion coupling [87]. Micromotion is an often unwanted part of the ion's motion and can be split into two forms; intrinsic and extrinsic micromotion. The latter is caused when the ion is not in the RF nil (caused by an external field contribution) and hence is subject to an additional, oscillatory component of motion. This can be compensated for by moving the ion to the RF nil. Intrinsic micromotion is the result of the ion's secular motion causing an effective offset of the RF nil in which the ion moves with an additional motion at the drive frequency *Ω*. This occurs even at the RF nil and cannot be compensated for entirely (see Ref. [88] for a detailed discussion of micromotion).

This Technical Review overviews the state-of-the-art of microfabricated ion traps including efforts to integrate advanced features such as optical components and electrical devices. For a further discussion of ion trap supporting hardware, and other ion trap fabrication methods, we suggest Refs [35]–[40].

# 2. Ion Trap Geometries

In this section we discuss basic ion trap geometries for top layer electrode design, and the geometrical considerations for ion transport and advanced ion trap designs. The basics of ion trapping are covered in Box 1 and a typical ion trap experimental setup is described in Box 2. The evolution of ion trap geometries is discussed in Box 3.

## 2.1 Basic principles of five-wire geometry

Panel c in Box 3 shows a simplified planar view of a surface ion trap in a symmetric, five-wire geometry. An RF voltage is applied to a pair of linear electrodes while all the other electrodes are held at RF ground. The ponderomotive potential generated by the RF voltage confines ions parallel to the *z* axis at a height given by the widths of the RF and central ground electrode. Assuming infinitely long rails, the zero potential line from the RF, the RF nil, can be expanded along the longitudinal direction, and the axial position of the ions can be determined by static electric fields only. To generate the static electric field required for trapping or shuttling the ions in the axial direction, a calculated, DC voltage set is applied to the segmented electrodes [54], [55].

The five-wire electrode geometry of surface ion traps has been analytically modelled in numerous studies [28], [56], [57] including the modelling of electrostatics due to gaps between electrodes [58]. The width of the RF and ground (denoted $a$ and $b$ respectively in panel c in Box 3) electrode in a gapless, five-wire geometry can be used to determine the trap depth, $\psi_E$, and ion height, $h_{\mathrm{RFnil}}$ Ref. [56];

*Equation 1*

$$\psi_E = \frac{e^2V^2}{\pi^2 m\Omega^2}\frac{b^2}{(a+b)^2+(a+b)\sqrt{2ab+a^2}},$$

*Equation 2*

$$h_{RFnil} = \frac{\sqrt{2ab+a^2}}{2},$$

where the trap depth is the difference in the ponderomotive potential between the RF nil and the escape point (that is the energy of the ion is larger than the trapping potential), and $\Omega$, $e$, $m$, and $V$ indicate the RF drive frequency in Hz, elementary charge in Coulombs, ion mass in kilograms, and RF voltage amplitude in volts, respectively.

A high trap depth is required to trap ions for a long time. For a low ion motional heating rate, a large ion-electrode distance is wanted. These parameters cannot be optimised simultaneously (especially given differing scaling laws), and compromises are determined by considering the constraints given by the specifics of the experimental setup [59], [60]. In Ref. [60] it was analytically demonstrated that a ratio of RF and ground rails widths of $b/a$ = 3.68, provided a maximised trap depth. However, this wide ratio, increases the distance between the outer DC electrodes and the trapped ions, thus reducing DC confinement. A solution to this, replaces the central ground electrode, with segmented DC electrodes allowing for a minimised ion-electrode distance [55], however, depending on fabrication constraints, this is not always achievable.  This balancing act is common place in ion trap design and can depend heavily on the purpose of the experiment and voltage range available on the electrodes. The longitudinal (axial) direction is not usually considered during simple, RF electrode design, since its properties are mainly determined by DC voltages.

## BOX 2: An ion trap experiment

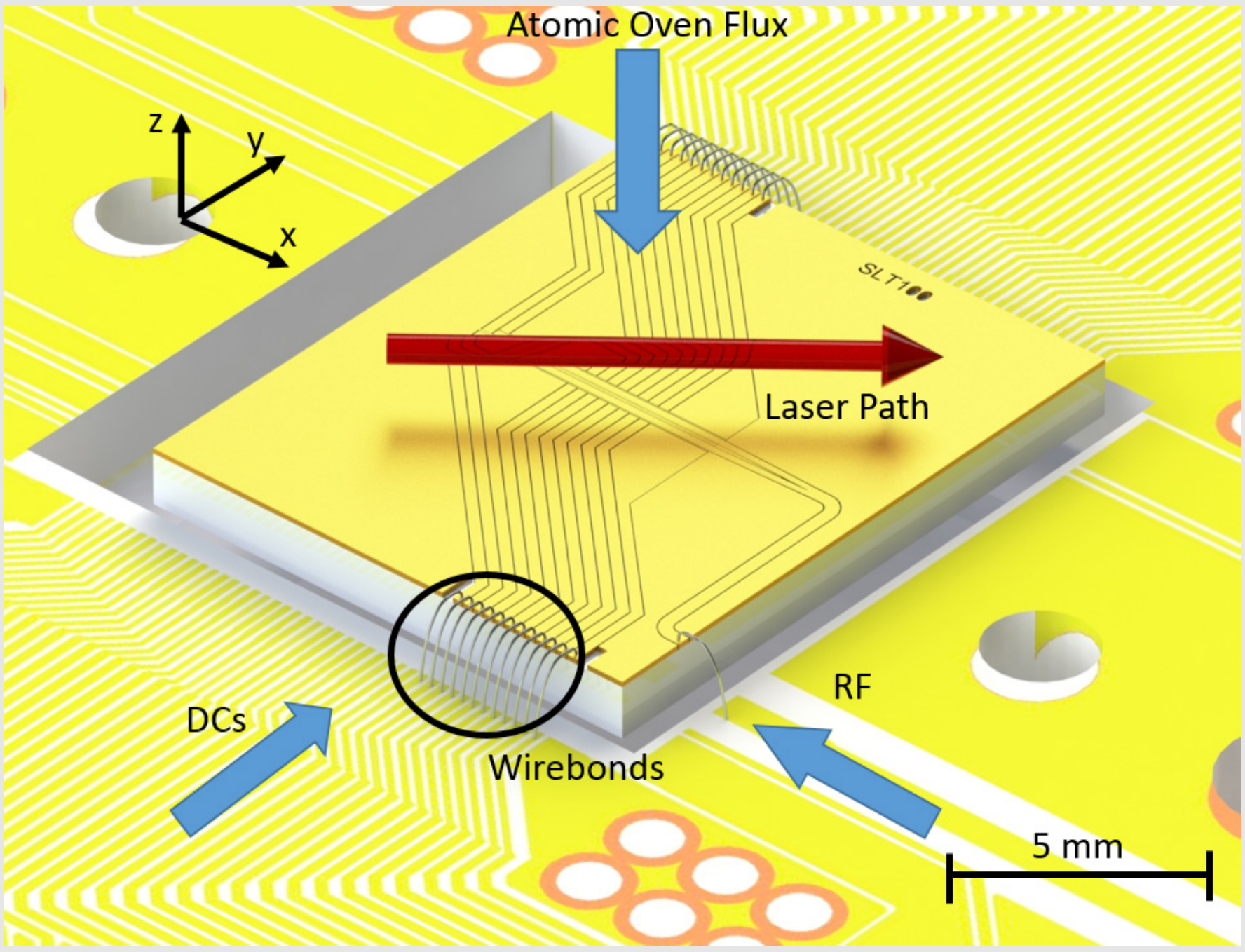

In a typical ion trap experiment setup the ions are trapped in an ultra-high vacuum (UHV) chamber with feedthroughs to connect the ion trap to supporting electronics and viewports for optical access. An atomic oven with a small aperture produces a flux of atoms parallel to the trap surface [89]; a laser ionises the atoms [90]. The atomic flux can lead to unwanted surface coating, and many efforts have been made to reduce this, for example by loading through a hole in the trap itself [51]. To initially cool the ions, Doppler cooling is used [91]. When Doppler cooling, all 3 directions of motion must be considered to effectively reduce the ion motion. To cool the in-plane directions of motion, the laser is typically positioned, parallel to the surface, at a 45° angle with respect to x and y directions of motion to have cooling components in both. The *z* motion can be effectively cooled by rotating the principal axis using the methods discussed in the main text. Additional lasers and laser paths may be required to perform laser-based quantum logic operations [92].

An important consideration for ion trapping is optical access to the ions. Wirebonds are often used to connect the trap to a supporting printed circuit board (PCB). If the wirebonds are in the path of a laser, unwanted scattering will occur, which will drastically affect the ability to perform experiments. Scattering also occurs due to the divergence of a planar laser beam, scattering off the surface of the trap. Some surface ion traps have been especially developed to reduce this scattering by including an optical access hole which allows laser access through the trap [62]. To image the ion, external optics, combined with charged-coupled devices (CCDs) and photo-multiplier tubes (PMTs) are used. These sensors and optics are typically outside the vacuum system, however efforts are being made to integrate the required technology for the detection of ions into the ion trap structure.

## 2.2 Rotation of principle axes

In a typical experimental setup with a surface ion trap, the allowed laser paths are limited to the directions parallel or perpendicular (through a vertical hole penetrating the substrate [55]) to the surface of the trap chip. To be able to effectively Doppler cool an ion in all motional directions (principal axes), the laser path must interact with all the principal axes. This is achieved by rotating the principal axes of motion (Fig. 2.1). The rotation can be achieved by tilting the total electric potential at the ion position. The rotation angle can be calculated from the eigenvectors of the Hessian matrix of the total electric potential. To tilt the potential, there are two commonly used methods. The first approach uses RF rails of different widths, which rotates the potential [23], [57], [61]. The second method uses asymmetric DC voltages applied to 'rotation' electrodes. Rotation electrodes can be introduced by replacing the central ground electrode in a 5-wire geometry with two electrodes which occupy the same space, creating a 6-wire geometry. As this will move the ion out of the RF nil, additional voltages are applied to the control electrodes (Fig. 2.1) and are required to compensate for non-zero fields [54], [62]. Asymmetric RF electrodes were initially used for surface traps, until proposals were made to rotate the principal axes using DC voltages instead. The use of asymmetric DC voltages applied to these electrodes, became a popular approach since it could easily achieve an 'out-of-plane' principal axis angle of rotation of θ = 45° , allowing for all degrees of freedom to be addressed by a planar laser path [54].

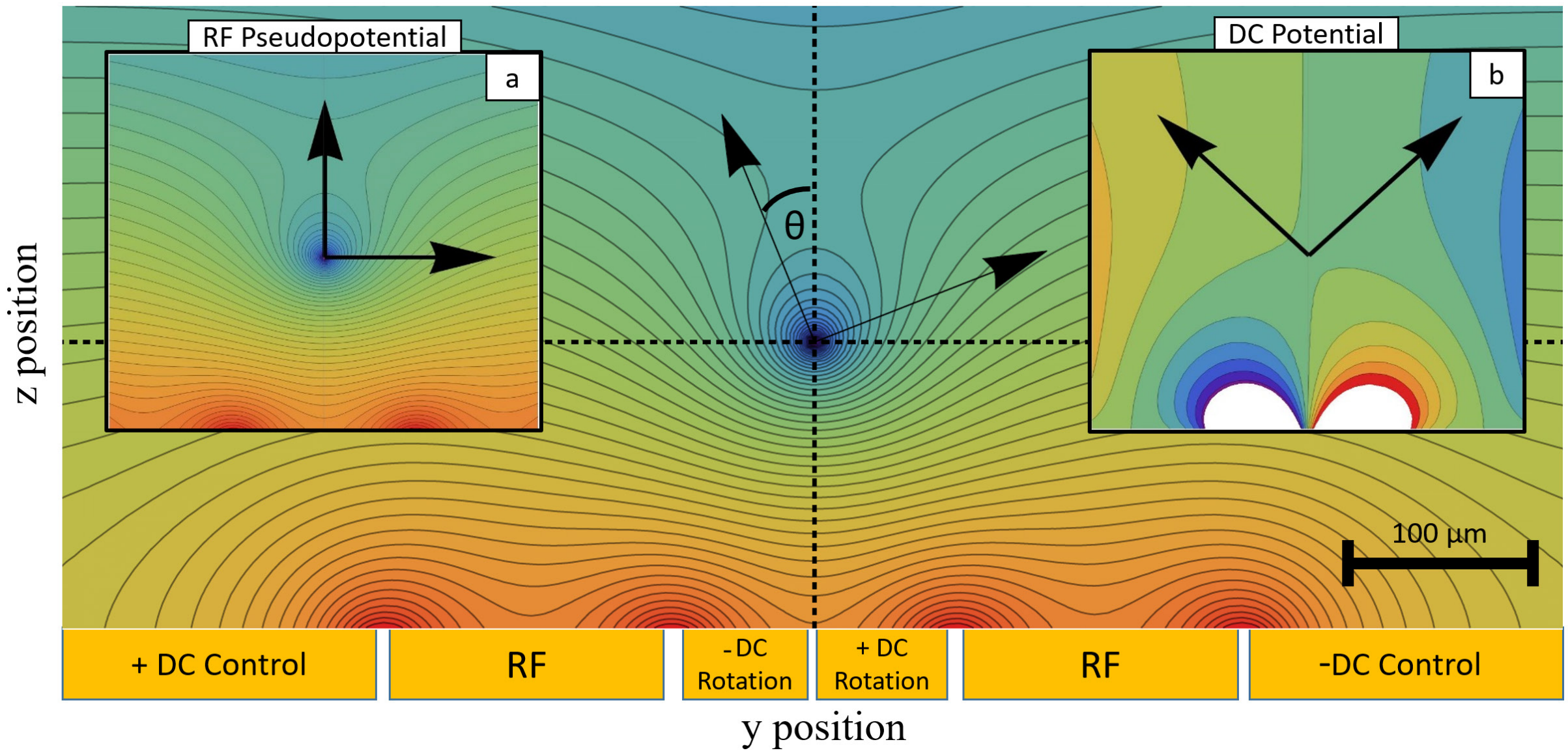

**Figure 2.1** Method of rotating the principal axis by angle, θ using a 6 wire surface trap design [54]. The arrows show the principal axis in the radial directions. The relative sign of the voltage to achieve the rotation is indicated on the electrodes. The scale indicates typical dimensions of the electrodes, however, these can vastly differ depending on the wanted ion height and trap parameters. The central contour plot shows the total potential, $\varphi_{tot}$ created from the superposition of **(a)** the RF Pseudopotential and **(b)** the DC rotation potential created by asymmetric voltages on DC electrodes.

## 2.3 Design and optimisation of geometries for ion transport and quantum simulation

Some applications of trapped ions, especially quantum computation, require the ability to move ions in a trapping potential such that certain operations are only performed on particular ions. In general, there are four types of operations required: linear shuttling, junction shuttling and separation and combination of ion crystals. These operations are carried out using time-dependent voltages applied to control electrodes which are located on the trap. The optimal geometries of these electrodes (in relation to electrode-ion distance) for operations such as linear shuttling and ion crystal (re)combination have been discussed in Refs [56], [60]. These transport operations have been reliably demonstrated with high fidelity [63]–[67], based on the theoretical work presented in Refs [68]–[70].

## BOX 3: Evolution of ion trap structures

The first Paul trap (panel a) used hyperbolic electrodes in a 3D configuration [1], and with this structure, the isolation of a single ion [41] and the demonstration of a quantum logic gate [42] were performed. However, this structure could only trap a single ion without significant micromotion, which limited the measurement accuracy of atomic resonances. To address this limitation, the linear Paul trap was developed [43], [44]. These traps consist of four machined rods assembled in parallel to confine ions radially and two end cap electrodes to axially confine ions [45] (panel b). One of the most important characteristics of this linear trap is that ions in the same string share their motional modes.

High voltages (>1 kV) can be applied to the four-rod trap, which allows the creation of deep trapping potentials whilst maintaining stable trapping parameters. The electrodes are sufficiently far from the ions to minimise the effects of electrical noise from the electrodes. Segmenting these rods allows axial control of the trapping potential so these traps are used in experiments with ion chains in situations that do not require microfabricated devices [46]–[49].

Several quantum computing architectures [17], [19], [52], [53] require reliable, reproducible devices and precise control over the ion's position, the latter being realised with more electrodes of smaller size. To address this, lithographic techniques from the semiconductor industry were used to fabricate ion traps [22], [50]. These ion trap reproduced the four-rod structure on a micron-scale, but the vertical distance between the electrodes was inevitably limited by the technical capability of thin film processes. A solution was proposed in Ref. [28] and subsequently demonstrated in Ref. [23]: a direct projection of the four rods onto a single plane resulting in a five-wire geometry. This trap lays two RF electrodes and three ground electrodes alternately, as shown in panel c. Here 'a' and 'b' represent the widths of the ground and RF electrodes respectively and the yellow arrows indicate the electric field when the RF electrode has a positive voltage applied. This chip structure is referred to as a `surface electrode trap' (or `surface trap') and has become the most widely used geometry for microfabricated ion trap chips.

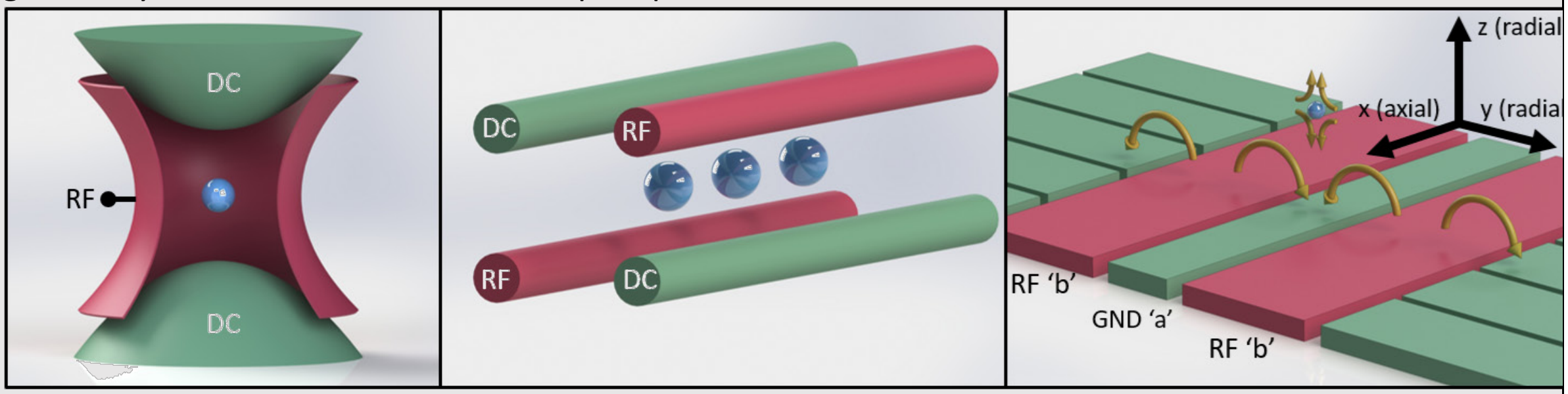

(a) (b) (c)

Advances in fabrication technologies have allowed more complex designs to be pursued [17],[26], [27], [29], [51].

Whilst linear shuttling operations can be performed using the types of ion traps previously discussed, junction shuttling, and by extension, ion trapping in 2D grids or planes, require several modifications. Arrays of trapping zones arranged in a 2D plane were first proposed in Ref. [17] and expanded towards and industrial blueprint for quantum computing in Ref. [19]. The first realisation of ion transport through a junction was demonstrated in a T-junction ion trap array [71]. Subsequent studies identified optimal geometries for ion trap arrays where linear regions are connected to others through junction nodes [72], [73]. At the centre of a junction node, three (T [71] or Y-junction [51]) or four (X-junction [74]) branches of linear rails join together, making the infinitely long rail assumption no longer valid. Consequently, the uniform extension of the RF nil along the axial direction terminates at some point (Fig. 2.2). Changes in the ion's secular frequencies or the transport over a gradient in the pseudopotential, such as that caused by a junction, can lead to the motional heating of the ion [74], [75]. To minimise these effects, optimised geometries have been introduced to improve junctions [73], [76], [77]. The majority of these optimised geometries are created using iterative optimisation methods such as a genetic algorithm. Using these designs, a number of successful experimental results of junction transport have been reported [51], [74], [78], achieving $10^5$ consecutive transports with Doppler cooling and 65 without [76]. Furthermore, Ref. [67] demonstrated the preservation of quantum information during trapped ion transport with 99.9994% state fidelity.

Whilst the previous designs are used for generic ion traps, for quantum simulations, 2D ion lattices of stationary ions (with small inter-site distances) can be beneficial. Ion traps to create these lattices have been successfully fabricated as mechanical structures [59], and subsequently as a microfabricated ion trap chip [26]. Ref. [79] introduced a useful tool for creating geometries required for close lattice sites. Using this tool, lattice geometries have been fabricated with close, inter-site distance and multiple degrees of freedom per site [27], [80]. This tool has also been used to investigate bi-layer ion traps which can be used to achieve stronger coupling between adjacent sites than with lattices fabricated on surface ion traps [81].

## 2.4 Numerical Simulation Tools

As the scope of ion trap experiments has evolved, so have the requirements of the ion traps themselves. Such requirements include the introduction of vertical holes penetrating the substrate for additional access of laser light or atomic flux to the ion position [82], or the split rotation rails and the junction geometries already discussed before. For those geometries, analytical methods are no longer viable, therefore numerical simulations of the electric fields are essential for designing electrodes. In the early days of surface traps, the boundary element method was used to simulate simple geometries with a single electrode layer [83], owing to the available computational resources. Advances in computational power meant that the finite element method become a viable simulation tool and geometries with greater complexity, including structures for oscillating magnetic field gradient schemes, are routinely modelled and optimised using this method [77], [84], [85]. The X-junction device in Fig. 2.2 has been optimised to reduce the pseudopotential gradient which, in turn, reduces the motional heating rate , $\dot{\bar{n}}$, during ion transport [76].

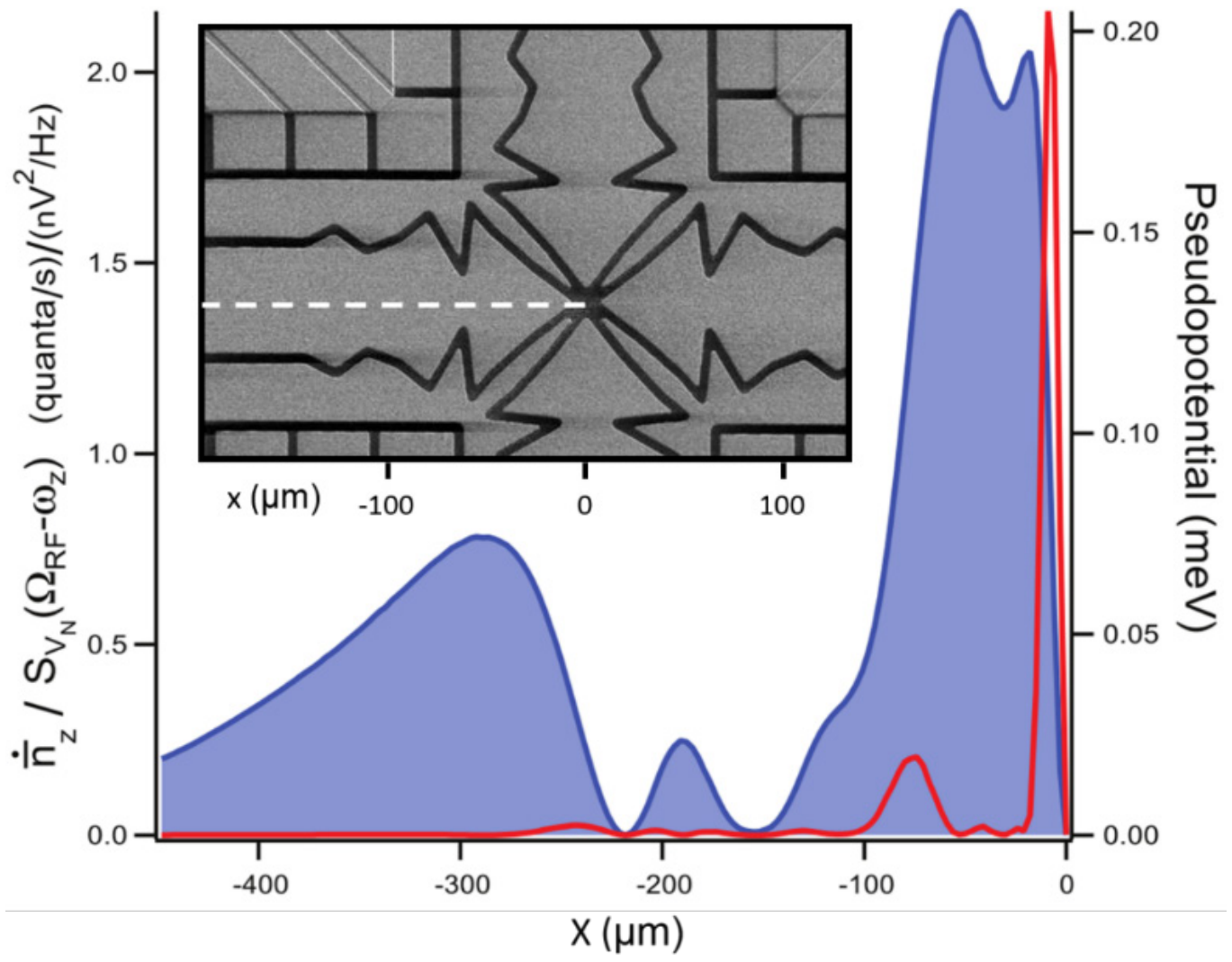

**Figure 2.2.** Optimised x-junction electrode geometry at the junction centre (inset image) to reduce the ratio of the motional heating, $\dot{n}$ , to the spectral voltage noise, $S_{V_N}$ (red line). It is expressed as the quantity $\dot{n}\ /S_{V_N}$ to normalise against material and electronics dependent noise. This ratio provides a measure of the gradient in the pseudopotential and local secular frequency both of which determine the motional heating during ion transport [74]. The blue shaded line shows the RF pseudopotential along the transport direction. The trap potential is evaluated for a $^{40}$Ca ion with V= 91 $V_{rms}$ and Ω= 2π x 58.55 MHz. Figure taken from [76] and modified for continuity.

# 3. Ion Trap Microfabrication Techniques

The semiconductor industry revolutionised the creation of miniature, electronic devices using patterned conducting and insulating materials, the most common example being CMOS. Key to this success are highly reliable and reproducible processes using well-established fabrication facilities (foundries). Microfabricated ion traps have been made in such foundries [30], [32]. However, some ion trap specific requirements that necessitate structures which do not use established processes (such as formation and subsequent patterning of thick dielectric layers) or materials (such as gold and copper) are not possible in many foundries. As a result, modern ion trap fabrication borrows techniques across different microfabrication technologies, such as MEMS, with the ultimate goal of achieving reproducible processes such as those used in CMOS.

## 3.1 General considerations for the microfabrication of ion trap chips

Critical features that should be considered when designing ion trap chip structures and the required fabrication processes are as follows:

- The electrodes should be able to sustain a RF voltage suitable for that particular ion species (heavier ions require larger voltages). A higher RF voltage allows trapping of ions farther from the chip surface, reducing the effects of the electrical field noise and laser scattering from the chip surface. In addition, higher voltages can allow for higher trap depths and secular frequencies.
- Coupling of the RF field into the substrate should be minimised to prevent power loss and heating of the device [72].
- The area of dielectric exposed to the trapped ions should be minimised. In general, any dielectric should be sufficiently shielded such that electric fields from trapped charges on the dielectric, are not felt by the ion. This is because factors such as ultraviolet (UV) lasers incident on dielectric surfaces can generate time-varying, stray charges in the dielectric [93]–[95], which in turn cause unwanted, time-dependent, ion displacements.

- All the materials, including various deposited films, should be compatible with ultra-high vacuum (UHV) environments. The chip should also be able to withstand processes required to achieve UHV such as baking and cryogenic temperatures.
- There must be unobstructed optical access between the lasers sources, detectors and ions. This requires, for example, that wirebonds do not impede the optical path.
- Exposed surfaces should be contaminant free to the greatest extent to reduce the anomalous heating of ions [96], [97]. In addition, a lower surface roughness reduces unwanted laser scattering and has been suggested to also reduce anomalous heating at cryogenic temperatures [98].

## 3.2 Substrate materials

Designing a fabrication process for ion trap chips starts with the selection of layer materials, since the number and dimensions of thin film layers can drastically change the complexity of the entire process. Dielectric and conductive (or semi-conductive) substrates have their own distinct advantages and disadvantages. The use of dielectric substrates allows for a very simple fabrication process and low power loss and heat in the substrate. However, accurate bulk micromachining of dielectric substrates for introducing vertical penetration holes and buried metal layers can be difficult. Another concern is that of exposed dielectric surfaces can trap charges, causing stray electric fields [93].

Silicon is the most widely used material in modern semiconductor technology and has the advantage of using most processes that are not compatible with insulating substrates. It is, however, very lossy in the mid-range resistivity ($10^2$-$10^4$ Ω cm) for RF frequencies [99]. To compensate this, the simplest and most widely used method is to place an additional metal layer referred to as a ‘ground plane’ (M1, Fig. 3.1(a)) between the RF electrode and the substrate. Although this adds complexities to the fabrication of silicon ion trap chips, it is widely adopted since it almost guarantees the successful shielding of the substrate from RF dissipation. Another approach is to use either very high or very low resistive substrates [100]. In this case, the temperature dependence of the silicon resistivity should be considered: at cryogenic temperatures, silicon can even act as an insulator [101].

## 3.3 Standard fabrication processes for ion traps

A typical fabrication process of a silicon based, surface ion trap chip is shown in Fig. 3.1. The trap consists of a ground layer (M1), an electrode layer (M2), and two dielectric layers (D1 and D2) that are used to electrically isolate the conducting layers and the substrate. When using an insulating substrate, a simple ion trap can be created using M2 only and its formation is similar to that of M2 described in this section. The process starts with the deposition of a dielectric layer on the silicon substrate, which insulates the ground plane from the substrate. A 1-2 µm layer of either $SiO_2$ or $SiN_x$ is used for this and is typically grown using plasma enhanced chemical vapour deposition (PECVD) or as a thermal oxide. A ground plane, M1, is deposited on the D1 layer (Fig. 3.1(a)) using any UHV compatible conductive material such as Au, Al, Cu, and so on. This is commonly a 1-2 µm layer, either sputtered or evaporated onto the device. When developing chips with vertical interconnect access (vias) to improve electrode routing, these thin insulating and conducting layers can be stacked multiple times [102]. Since high-RF voltages applied between the electrodes and ground plane are desirable in surface ion traps, the D2 layer separating the two metal layers should be thick to maximise the voltage at which electrical breakdown occurs. Thus, the thickness of the D2 layer is generally on the order of 10 µm, with the deposition and etching of this layer usually being the most difficult steps throughout the entire ion trap fabrication process (Fig. 3.1(b)). This is because such vertical dimensions are generally not covered by conventional semiconductor processes, which also limits the material choice for the dielectric layer to plasma enhanced chemical vapour deposition $SiO_2$ or $SiN_x$. Ref. [103] demonstrated a polymer-based, spin-on dielectric for ion trap fabrication, which allowed for thick dielectric layers. The deposition of the top metal layer (M2) can be performed by using conventional microfabrication processes including electroplating, sputtering, evaporation (Fig. 3.1(c)) and can be patterned using plasma dry-etching (Fig. 3.1(d)). Since the top layer is directly exposed to the trapped ions, more factors should be considered when selecting the electrode material. Gold is one of the most widely used materials for the top metal layer, owing to its extremely low oxidisation rate. However, since gold is not compatible with many conventional microfabrication techniques, a thin gold layer can be deposited on aluminium electrodes as a final step [104]. The final step etches the dielectric layer in both the vertical and lateral directions to reduce the effect of charges trapped on the dielectric surface. The vertical etching of the thick dielectric layer (Fig. 3.1(e)) uses the electrode pattern as a mask, and reactive ion etch to etch. This is followed by isotropic wet or gas etching of the dielectric sidewalls (Fig. 3.1(f)), which helps to minimise the exposure of the ions to trapped charges [62].

The fabrication methods described here are commonly used to create a silicon-based ion trap chip. Microfabricated ion traps can also be made by many different, additional processes not discussed in this Technical Review [35], [38]. Furthermore, fabrication processes have been optimised for specific capabilities, such as extremely high breakdown voltage [26], [105] or entire shielding of dielectric sidewalls [95]. 3D quadrupole traps have also been fabricated using new techniques which allow for large ion-electrode distances (hence low motional heating rate) for a lower trap drive voltage than surface ion traps [29], [106]. Ref. [107] used a transparent material, indium tin oxide (ITO), as the electrode layer to detect fluorescence emitted from ions with a photodetector underneath the electrode on the backside of the trap chip. A selection of various microfabricated ion traps, which includes both surface and 3D quadrupole traps, are shown in Fig. 3.2.

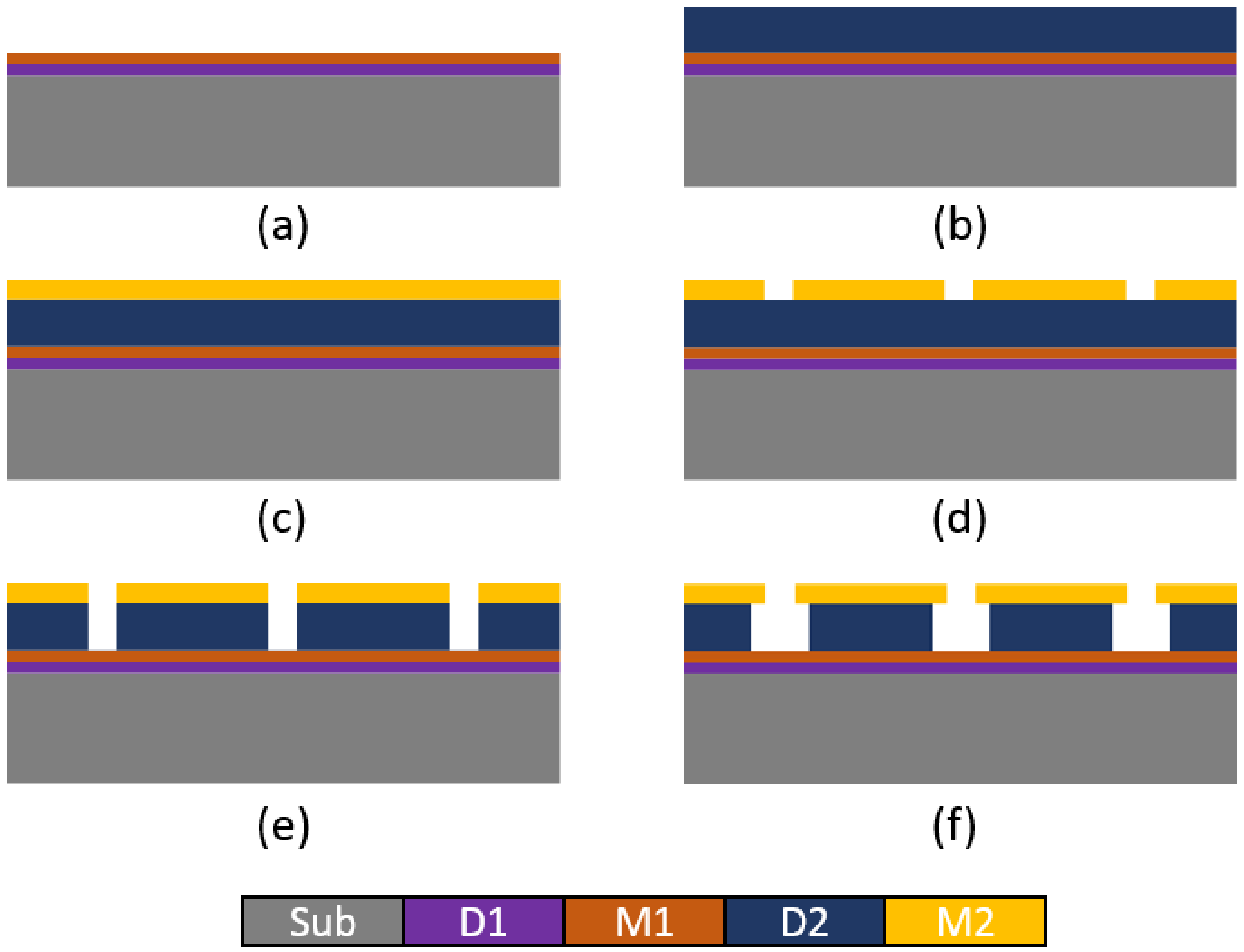

**Figure 3.1.** A conventional fabrication process flow of a surface ion trap chip. **(a)** Forming a ground plane and an insulating layer that isolates the ground plane and the substrate. **(b)** Deposition of a thick (~µm's) dielectric layer. **(c)** Forming a metal layer on the dielectric. **(d)** Etching of the metal layer to define the electrode patterns. **(e)** Subsequent etching of the thick dielectric layer. **(f)** Isotropic etching of the dielectric pillars from the sidewalls to reduce the area of dielectric sidewalls exposed to the ion.

In order to integrate various on-chip features, post or pre-processing steps can be added to the fabrication flow of the basic ion trap structures described previously. As an example of post processing, a vertical slot penetrating the silicon substrate can be fabricated using a conventional deep silicon etching process at the end of ion trap fabrication. This hole can be used to load neutral atoms [109] or to provide increased optical access [110]. For the integration of optical or electrical devices the ion trap is fabricated on top of pre-processed wafers where the integrated feature has already been fabricated. This inevitably introduces restrictions on the processes and materials used, hence these concerns should be addressed to provide cross-compatibility between fabrications. For ion traps which use multiple integrated technologies, the compatibility between processes can be an extremely difficult problem to solve and will likely become a strong focus of the ion trap community.

The fabrication methods described here are commonly used to implement a silicon-based ion trap chip. Microfabricated ion traps can also be fabricated by many different, additional processes not discussed in this review [35], [38]. Furthermore, fabrication processes have been optimised for specific capabilities, such as extremely high breakdown voltage [26], [105] or entire shielding of dielectric sidewalls [95]. 3D quadrupole traps have also been fabricated using novel techniques which allow for large ion-electrode distances (hence low motional heating rate) for a lower trap drive voltage than surface ion traps [29], [106]. Eltony *et al.* [107] used a transparent material, indium tin oxide (ITO), as the electrode layer to detect fluorescence emitted from ions with a photodetector underneath the electrode on the backside of the trap chip. A selection of various microfabricated ion traps, which includes both surface and 3D quadrupole traps, are shown in Fig.3.2.

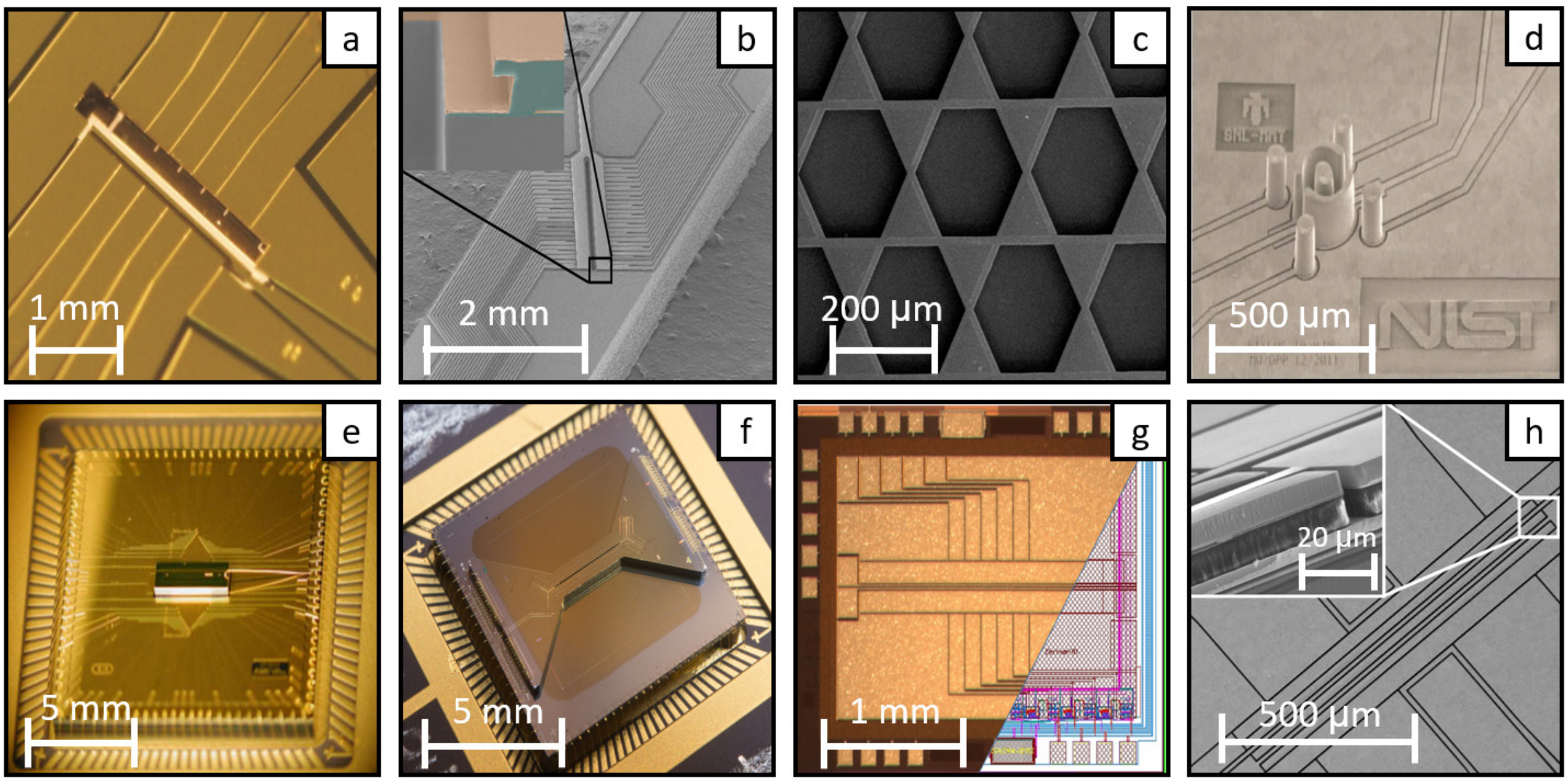

**Figure 3.2** – Images of reported ion traps **(a)** Microfabricated, 3D quadrupole trap **-** National Physics Laboratory (NPL) - Wilpers *et al.* [29] **(b)** Low dielectric exposure - Seoul National University – Hong *et al.* [95] **(c)** First 2-D array on a chip and high breakdown voltage - University of Sussex – Sterling *et al.* [26] **(d)** Large metal structures – National Institute of Standards and Technology (NIST) – Arrington *et al.* [108] **(e)** Through Silicon Via (TSV) in an ion trap - GTRI/Honeywell – Guise *et al.* [102] **(f)** High optical access – Sandia National Laboratory (SNL) – Moehring *et al.*/ Maunz [51], [55] **(g)** Ion trap fabricated in a CMOS foundry - Massachusetts Institute of Technology (MIT) – Stuart *et al.* [32] **(h)** Novel fabrication method for thick metal/dielectric layers - Physikalisch-Technische Bundesanstalt (PTB) – Bautista-Salvador *et al.* [103].

## BOX 3: Efforts to deal with heating rates

Despite many successful implementations of surface traps, miniaturising the ion trap structure has also created a number of side effects, one being the so-called `anomalous heating' of the trapped ions [21]. This is thought to be induced by electric field noise from the surface of the trap electrodes. The electric field noise can couple to the ion motion when the frequency is near the motional frequency of the ion, which in turn leads to an increase in the phonon number which is detrimental to the implementation of quantum logic operations. Many efforts have been made to investigate this heating both theoretically [111], [112] and experimentally [113]–[115]. One method of reducing the heating rate is by increasing the distance, *d*, between the ions and the electrode surface, since it has been experimentally shown that the heating rate scales as ~$d^{-4}$ (Refs [114], [116], [117]). However, given that the maximum voltage that can be applied to a trap chip is limited by current semiconductor technologies, an increased distance inevitably leads to a shallower trap depth, therefore, there is a limit to how much *d* can be increased. Another method to reduce the heating rate is by cooling the ion trap to cryogenic temperatures, first demonstrated in Ref. [114]. A number of publications showed that this approach can reduce the heating rate by two orders of magnitude [114], [61], [118]. Other approaches use in-situ cleaning of the chip surface [96], [119]. Since hydrocarbon-based contaminants adsorbed on the electrode surface during the bake-out process are suspected to be a major source of the electrical noise, inducing anomalous heating [120], the removal of these contaminants after the bake-out can also reduce the heating rate by two orders of magnitude. Owing to these efforts, extremely low heating rates of ions have been reported [100]. However, more work is required and is currently under way to better understand the source of anomalous heating [121]–[126].

The figure below summarises the heating rates in microfabricated ion traps as a function of ion-electrode distance. The measurement results of heating rates are expressed as noise spectral density with a fit line demonstrating a heating rate scaling of $d^{-3.79}$ [117]. The heating rates measured in a cryogenic environment are noticeably lower than those in room temperature systems. The red and blue arrows indicate the changes made in the same experimental setup by in-situ surface cleaning and cryogenic cooling, respectively. The inset shows the temperature dependence of the heating rate, showing significant gains from room temperature to cryogenic temperatures. Inset reproduced with permission from Ref. [118].

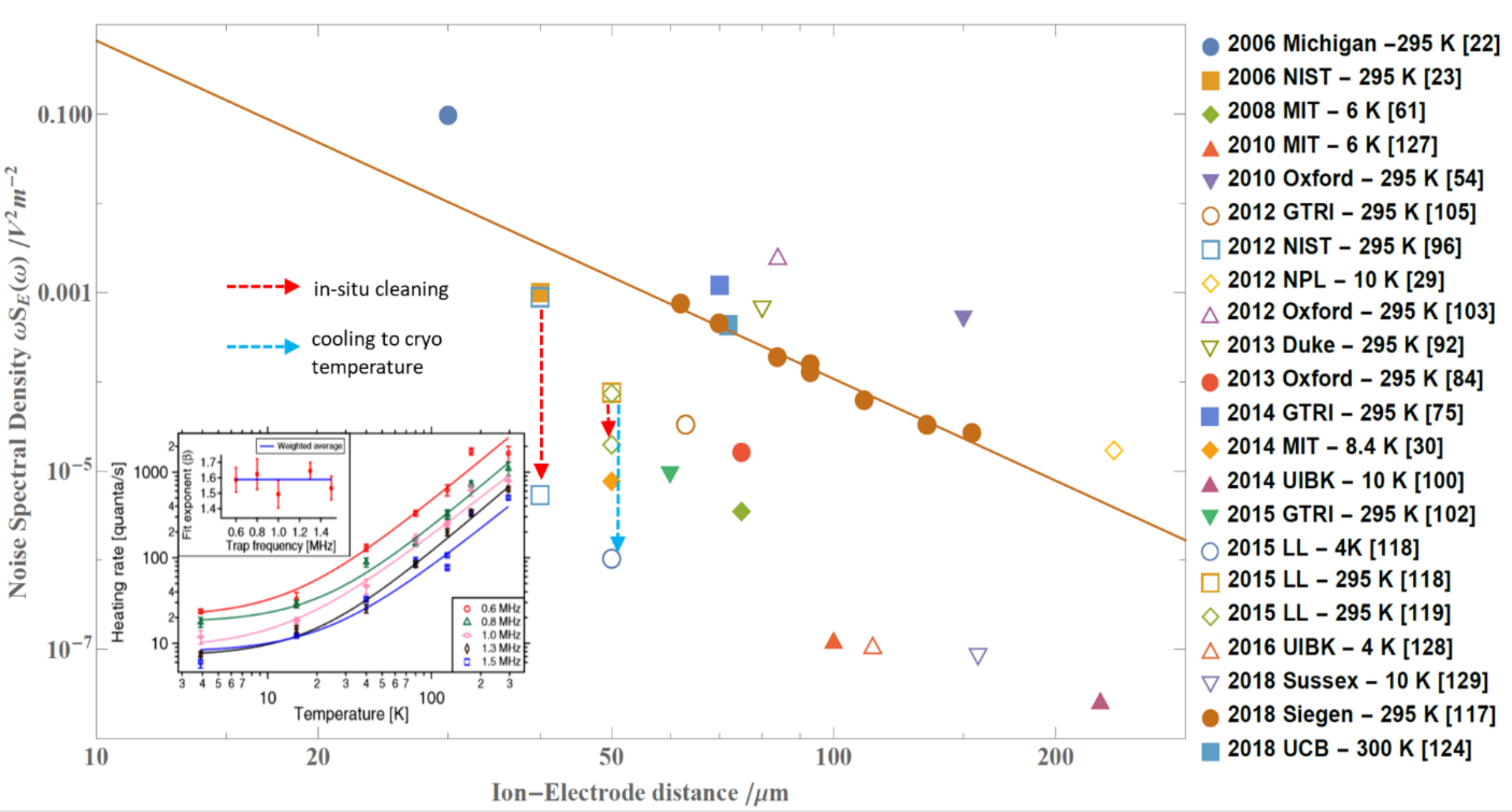

# 4. Advanced On-Chip Features

As ion trap geometries become more complex, either for quantum simulation or scalable quantum computation purposes, supporting systems are often required to sit within the footprint of the ion trap device [19]. Examples include integrating optical and electrical devices used in experiments, into the microchip. These approaches can also contribute to the miniaturisation of precise measurement devices such as atomic clocks and mass spectrometers, combined with efforts to miniaturise vacuum chambers [24], [34].This section mainly discuss research into chip-level integration of advanced features. Several of the processes for advanced feature integration use fabrication methods or materials that are not covered in this Technical Review. Readers are invited to the refer to the referenced material to learn about the extensive and varied processes as required.

## 4.1 Embedded Gate Schemes

Several schemes for implementing quantum logic gates between qubits rely on magnetic field gradients, either static [130], [131] or oscillating [132], [133]. Three methods to create the gradient have been used: permanent magnets, placed near the trap to produce a static gradient at the ion position [131], [134]–[136]; in-plane current carrying wires (CCWs), fabricated as part of the electrode layer, which allow the creation of both static and oscillating magnetic field gradients [84], [137]–[139]; sub-surface CCWs, fabricated below the electrode and ground plane layers [140].

Permanent magnets are commonly used for the static gradient scheme, achieving gradients of up to 36 $Tm^{-1}$ (Ref. [136]). Scaling up to large systems may prove difficult due to the need for a careful, manual alignment of the ion trap to the magnets.

In-plane CCWs have been shown to provide oscillating gradients of up to 54.8 $Tm^{-1}$ (Ref. [85]). In this scheme the trap geometry must be altered in order to accommodate the CCWs and the power dissipation must be considered. The power dissipation could be reduced by using thicker layers, to decrease resistance, however the skin depth of the oscillating field in the electrodes may quickly limit this, depending on the transition required and electrode material.

Sub-surface CCWs, situated beneath the ion trap, do not impinge on the electrode design and can be used for both static and oscillating gradients. For the latter, the field is likely to be attenuated due to shielding induced by currents in the ion trap electrodes and ground plane layers. The device in Ref. [140], used Cu wires 127 µm thick, fixed to an AlN chip carrier and mounted 285 µm below the trap surface. A gradient of 16 $Tm^{-1}$ was achieved using a current of 8.4 A.

Larger gradients could be achieved with sub-surface CCWs by reducing the wire to ion distance. Ref.[19] suggests that the CCWs could be imbedded into the substrate surface, using the dual damascene process, and the ion trap fabricated directly above the wires. By embedding the CCWs into the substrate, not only is the trap to wire distance decreased, but thermal sinking is improved, thus greater current, and therefore higher gradients, should be feasible [19]. Our group has already demonstrated the application of a current of 11 A (corresponding to a current density of $10^6$ A $cm^{-2}$) to an ion microchip which should result in a magnetic field gradient of >185 $Tm^{-1}$ at an ion height of 125 µm. At an ion height of 40 µm, this method is expected to produce a gradient in excess of 1,000 $Tm^{-1}$ or conversely obtain ~185 $Tm^{-1}$ with significantly lower current required (~3A).

Whilst a relatively new technology for ion traps, CCWs are common place in the atom-trapping community, where large, steep magnetic fields are used to trap neutral systems [141]–[144]. The currents and wire dimensions used for atom trapping are similar to those required by trapped ion gate schemes and therefore much of what has been learned in the atom trapping field may also be applicable.

## 4.2 Optical Components

For many trapped-ion technologies addressing and state readout of ions is required using optical techniques on an individual ion basis. It is sometimes beneficial that such optical components are integrated in the device, to increase the fidelity of addressing and readout operations.

Optical fibres have many uses in trapped-ion experiments for both addressing and readout operations. The typical structure of a fibre, however, is not naturally suited to an ion trapping experiment as an exposed dielectric near the ion can significantly disturb the trapping field. By integrating an optical access hole, the dielectric can be shielded by the ion trap itself, allowing for fibres to be brought close to the ion [82], [110], [145].

To address multiple ions, integrating optical waveguides into the ion trap is a promising way forward. Refs [146], [147] demonstrated a multi-ion addressing technique with integrated silicon nitride waveguides and grating couplers to address individual ions, with a total optical system loss of 33 dB. Using integrated waveguides [148], light of multiple wavelengths was delivered to a trapped ion. This demonstration was combined with packaging methods which allow for a direct optical fibre attach to the waveguides. Using these methods, a single-qubit gate with a $^{88}Sr^{+}$ ion was performed with 99% fidelity [147] and, more recently, two-qubit gates with $^{40}Ca^{+}$ were demonstrated [149]. Another approach for the scalable laser addressing is adjusting beam paths with electrically controlled devices. Ref. [150] demonstrated controlled beam steering with MEMS mirrors, however this technology has yet to be integrated into a trap.

Integrating mirrors into the ion trap surface can allow for more photons from the ion to be collected, by reflecting otherwise lost photons. In Ref. [151] micro-mirrors were integrated into a surface ion trap, enhancing photon collection of $^{40}Ca^{+}$ ion by 90% which resulted in a collection efficiency of 14% (numerical aperture of 0.69). The experiment in Ref. [152] showed a 4.1(6)% coupling of the fluorescence from a $^{174}Yb^{+}$ ion into a single mode fibre using integrated diffractive mirrors, which nearly tripled the bulk optics efficiency. Integrated mirrors can also be used to create cavities on chip to, for instance, facilitate atom-light coupling for photonic interconnects. Towards this endeavour, ion traps fabricated on top of high-finesse optical mirrors to create a cavity have been demonstrated [153], [154]. However, more development is needed before strong coupling, already demonstrated macroscopically [155], can be achieved in such a microfabricated device.

Standard ion trapping detection uses large collection optics which helps counteract the effect of being outside the vacuum system and hence farther from the ion [156]. By integrating a detector into the actual trap chip, the detector-ion distance decreases, which could help capturing more photons with appropriate optics in place. The studies in Refs [157] and [19] suggested using integrated single photon avalanche diodes (SPADs) for light detection. Ref. [107] presented a transparent trap made of ITO with an integrated photodetector featuring a collection efficiency approaching 50%. Another work [31] showed UV-sensitive superconducting nanowire single photon detectors (SNSPD) made of MoSi, integrated into a microfabricated ion trap. This device demonstrated a detector fidelity of 76(4)% at a wavelength of 315 nm with a background count rate below 1 count per second. The trapping field decreased the system detection efficiency by 9%, but did not increase background count rates. Being a superconducting device however, the stringent requirement on the operating temperature (3.2 K) introduces additional challenges, depending on the application. The use of SPADs, however, only requires temperatures of 70 K in order to achieve a performance similar to a photo-multiplier tube [19].

## 4.3 Passive Components

Capacitors, resistors and inductors are a key part of any ion trapping experiment in filtering, resonators and general electronics [24], [104], [158], [159]. Bringing these devices on-chip can have many benefits, from increased density of components to reduced noise due to the proximity to the device.

One method for the high-density integration of passive components takes advantage of advanced CMOS facilities by incorporating a CMOS die. Ref. [33] first showed this by integrating a 12 channel, bare-die RC filter array (35 kΩ, 220 pF) on to a printed circuit board. To attach the die, a standard low-outgassing silver epoxy was used, and channels were connected using wirebonds. Whilst not integrated on chip, the attachment methods are similar to those used by industry on silicon dies, hence, dies could be integrated on an ion trap if required. Using methods available at modern CMOS facilities [32] and careful calibration, one could realistically fabricate resistive temperature probes in the trap. Conversely, a resistive strip could also act as a local heater for use in cryogenic systems to prevent gas molecules freezing out on the trap during cool down [100].

Capacitors can be fabricated into the trap with two metal planes separated by a dielectric. Trench capacitors take advantage of an increased surface area by etching vertically into silicon using well established processes. This allows for a substantial increase in the capacitance per unit area. For ion traps, this was first in Ref. [104], where trench capacitors were integrated into the trap, which allowed 1 nF to be achieved in a 100 µm square. This was 30 times higher than the capacitance allowed in the same area when using a conventional, planar, fabrication process. A capacitive divider is a common method for measuring the large RF voltage applied to the ion trap [160]. Such a component could be integrated into the device itself using the methods discussed previously.

To create the low-noise, high-voltage RF trapping field, a helical resonator is commonly used [161], [162]. This component relies on the ability to impedance match using inductors. Efforts have been made to reduce the bulky nature of the device to a more manageable size [163], [164], but integration is yet to be achieved. Microfabricated inductors [165] could be one route forward in this respect. For future devices, microfabricated inductors could also be used to replace the standard low-pass filter [104] with more advanced filters, such as a band-stop filter, to remove noise on DC electrodes.

## 4.4 Active Components

In analogy to Rent's rule [166], it was suggested that for quantum computation, the scaling of interconnections and control lines with the number of qubits would become a major bottleneck [167]. This connectivity problem naturally lead to the introduction of active components into the vacuum system [33], or even into the ion trap [32]. One of the key requirements for ion trapping is that all integrated components must be UHV compatible [168]. This often eliminates the use of packaged electronics, making bare-die an UHV-compatible alternative. Packaged components can sometimes be used when operating at cryogenic temperatures. Ref. [33] introduced two, 40 channel DACs (AD5370) into an UHV environment. For this experiment, the AD5370 was not sourced in bare-die form, but instead a standard, packaged, AD5370 was used and then decapsulated using nitric and sulphuric acid [169]. The assembly is as discussed in the next section. Whilst this is a new method to overcome the low supply of bare-die products, it is likely that future devices using this technology would already be in bare-die form, hence suitable for UHV.

Ref. [32] integrated a custom, CMOS DAC into an ion trap. The custom DAC was designed for the CMHV7SF 180 nm node from Global Foundries. Key to this node is the ability to allow for higher voltages (20 V span) compared to typical <5 V span, thus allowing the implementation of an amplifier for voltages more suited for ion trapping. A switching device was also included to disconnect the ion trap from the DAC when not updating, hence reducing noise when disconnected. Since the electrode acts as a capacitor with low leakage current, the switch can disconnect the DAC whilst the electrode holds a voltage. The ion trap on top was also fabricated as part of the CMOS process, similar to that used in Ref. [30]. A common concern with semiconductor devices is the 'freeze-out' at low temperatures, where the energy required to overcome a band-gap becomes too large. However, the previously mentioned device was operated at cryogenic (4 K) temperatures and more complicated devices, such as a field programmable gate array (FPGA), have also been shown to work at such temperatures [170].

### 4.5 Stacked Wafer Technology

It has been proposed that back-side connections to the ion trap will be required for a large-scale quantum computer to connect to its supporting systems [102]. Ref. [19] expanded this to use the wafer stacking of different components, such as DACs, detectors, and cooling. These proposals require the introduction of through-substrate-via technology, often known as through-silicon-vias (TSV). The first demonstration of these technologies in ion traps was reported in Ref. [102], who connected an ion trap through a backside ball-grid array and TSVs. In Ref. [102], the ion trap die was attached using a ball-bonding method, which uses a programmed gold ball bonder to leave individual short, gold studs on an interposer. A localised eutectic bond was then used to connect the interposer to the back-side of the device. For a wafer-scale bonding, issues occur with high stress (due to the device size), reducing the connection quality [171]. On a wafer-scale, the studs are commonly microfabricated as opposed to using a ball bonder [172]. Eutectic bonding is also not the only option for wafer-scale attaches and many other, commonly used, methods exist [173], [174].

Ref. [19] suggests that the micro-channel cooling could be implemented as part of the stacked wafer proposal. This could also be used in other architectures, either for power dissipation or to reduce the heating rates. Whilst they have yet to be introduced into ion traps, micro-channel coolers have started to emerge for room temperature devices capable of removing >300 W/cm$^2$ (Ref. [175]). Ref. [176] demonstrated a micro-channel cooler using liquid nitrogen capable of removing 1000 W/cm$^2$, which was considerably helped by an order-of-magnitude increase in thermal conductivity of Silicon at ~70 K (Ref. [177]). Difficulties with this design, however, may emerge from the heat-flow between the stacked wafer levels and work is still required to develop heat-flow mechanisms between the wafers.

## 5. Outlook

Much progress has been made in the last 10 years on the development of complex microfabricated ion traps. Whilst the creation of custom potentials through simulations is well understood, many unanswered questions still remain in terms of the integration of advanced technologies. A summary of the requirements for many different advanced ion traps is illustrated in Fig. 5.1. With most of these technologies having already been individually demonstrated in trapped ion experiments, the microfabricated ion trap is expected to become a fundamental component for many quantum technologies.

Commercial realisation of quantum technologies requires the large-scale fabrication of devices, with high reliability and yields. Many of the proposed technologies discussed in this Technical Review will require CMOS devices integrated into the ion trap. However, unlike the standard operation environment of CMOS, ion traps are often required to work at cryogenic temperatures. Whereas the development of cryogenic control electronics is also a priority for superconducting qubit platforms [178] and solid-state qubits [179], temperature requirements for ion trap devices are much less stringent only requiring operation at 70K or possibly 4K where high cooling powers are available. Cryogenic CMOS integration of a DAC has recently been demonstrated in Ref. [32] which already has the capability to be mass produced. Whilst commercialised SNSPDs are relatively uncommon, SPADs are becoming increasingly wide-spread, with a high demand from the driverless car market. SNSPDs may prove simpler to integrate as they can be fabricated as part of the ion trap, however, they have to be operated at superconducting temperatures. Waveguide-on-chip technology is a developing field with many applications across the spectrum of quantum technologies. As a result, several fabrication facilities have appeared in the past few years which can offer some of the commercial capabilities required for the optical components discussed. However, further development is still needed in this field to create optical components with greater suitability for blue and near-UV wavelengths and introducing such components to an industrial-scale process. Wafer stacking and TSVs, whilst nascent technologies, are becoming prevalent in many different modern devices, such as high-bandwidth memory [180], and is a well-understood process [181]. Cooling requirements will become increasingly important to consider when integrating different technologies. Microchannel coolers provide a promising path towards managing the potential future thermal requirements however they have yet to reach market.

Although fabrication and assembly industries allow individual aspects of an ion trap to be demonstrated, combining these technologies will be essential for creating a more advanced ion traps. Separating the technologies on a wafer-by-wafer level may prove prudent in reducing risks of compatibility, but introduces new issues with inter-wafer connectivity for large devices and heat flow. Whilst integrating electronics can reduce some of the noise thanks to the proximity of the supporting electronics and hence less 'pick-up', it can introduce new noise spectra due to the integration of active components. This can be mitigated by additional filtering and noise manipulation to move the noise spectrum to frequencies that do not affect the ion. Some of the optical technologies discussed in this Technical Review will perform optimally only at certain wavelengths. Integrating many of these optical components could therefore require multi-species schemes to best take advantage of these technologies [182]–[186]. With a large variety of technologies being simultaneously integrated, unforeseen consequences will be a likely occurrence, hence full integration will remain a tremendous engineering challenge.

# 5. Conclusion

Trapped ions constitute an extremely powerful system to realise and control quantum phenomena such as entanglement and superposition, showcasing unmatched fidelities and coherence times. The emergence of ion microchips allows us to incorporate the advantages of modern microfabrication and microprocessor advances to this system, giving rise to a fully scalable hardware platform for a wide range of quantum technologies.

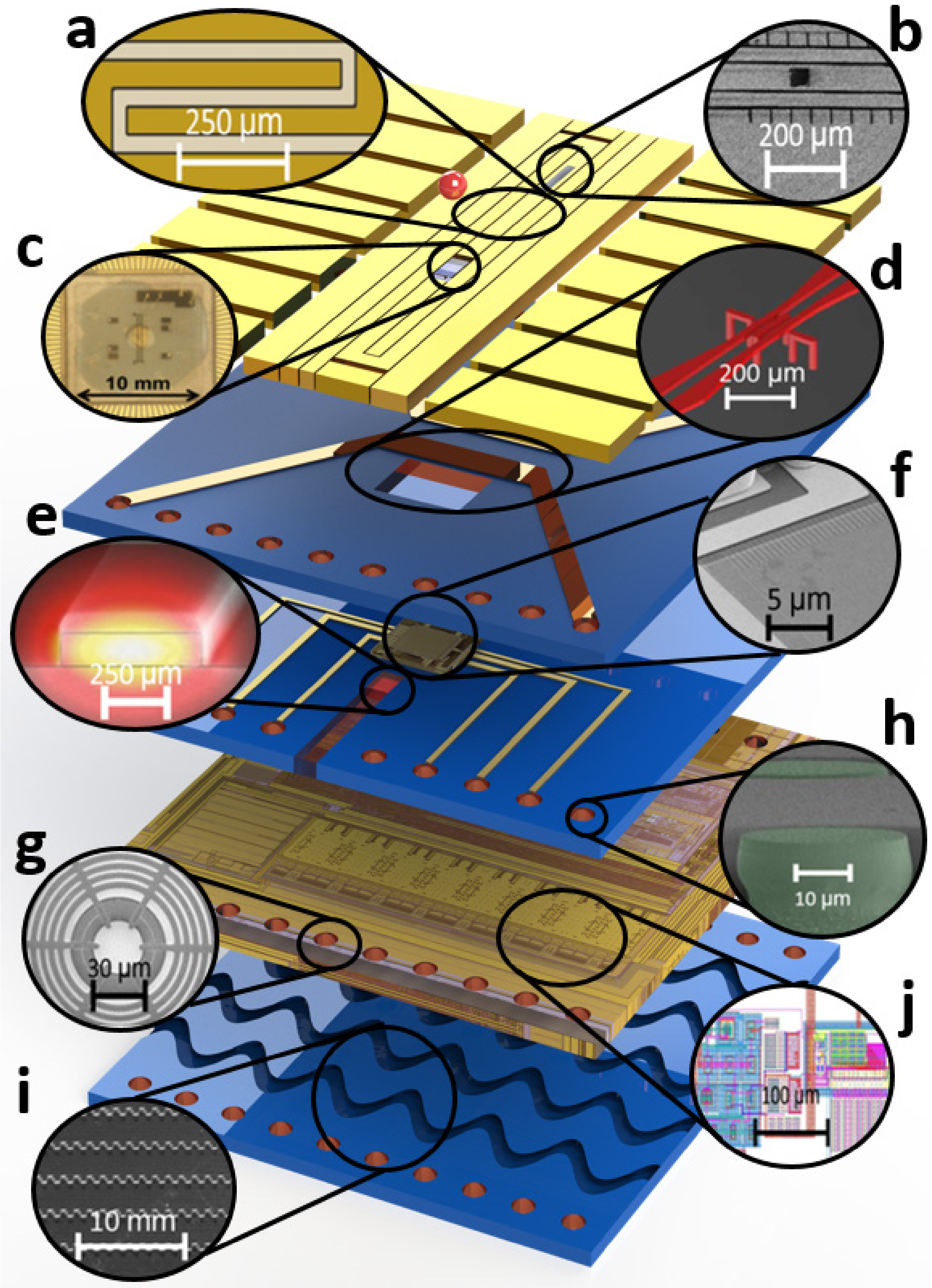

Figure 5.1 – Advanced on chip technology in an ion trap. It should be noted that not all features are required and their necessity is dependent upon use case. **(a)** Oscillating gradient CCWs [138] **(b)** Backside loading [76] **(c)** Transparent ITO electrode [107] **(d)** Static gradient CCWs [19] **(e)** $Si_3N_4$ waveguide and grating for individual optical addressing [147] **(f)** Integrated photon detector [31] **(g)** Trench capacitors [102] **(h)** Through Silicon Vias (TSVs) [102] **(i)** Microchannel cooling [175] **(j)** Integrated electronics [32].

# 6. Acknowledgements

This work is supported by the U.K. Engineering and Physical Sciences Research Council via the EPSRC Hub in Quantum Computing and Simulation (EP/T001062/1), the U.K. Quantum Technology hub for Networked Quantum Information Technologies (No. EP/M013243/1), the European Commission's Horizon-2020 Flagship on Quantum Technologies Project No. 820314 (MicroQC), the U.S. Army Research Office under Contract No. W911NF-14-2-0106, the Fonds National de la Recherche Luxembourg (National Research Fund) Project Code 11615035 and the University of Sussex.

## Glossary

Fidelity - The reliability of a certain operation. Fidelities above 99% for all operations can allow for the creation of a fault tolerant quantum computer [187].

RF nil (null) – The minimum energy of the RF pseudopotential and gives the ion’s position in an RF field.

MEMS – Micro-ElectroMechanical Systems are devices which combine electrical and mechanical features in a fabricated device. The sizes of the features used often make the processing similar to ion traps.

CMOS – Complementary Metal Oxide Semiconductors is a material structure which is used to make digital logic circuits. It is also a mature industry which produces highly reliable structures at high quantities.

Hessian matrix – Matrix of second order partial derivatives. In this case, derivatives of the total potential seen by the ion.

Pseudopotential – An effective potential which accounts for the ion’s motion in an oscillating electric field.

PECVD – Plasma Enhanced Chemical Vapour Deposition is a method of depositing materials through chemical reaction of ionised gases.

Sputtering – A method of depositing a variety of materials onto a surface using accelerated ion into a target of said material.

Evaporation – A method of depositing metals onto a surface by evaporation a metal and the resulting flux, coating the surface.

Reactive Ion Etch – Or plasma etching, a method that uses a plasma to directionally etch a material.

Breakdown – The point at which two, electrically isolated electrodes, become shorted due to a large voltage. This often damages the electrodes in the process.

CCWs – Current Carrying Wires are ion trap specific structures which are designed to use currents to generate magnetic fields.

Damascene process – A fabrication process in which a pattern is etched that is subsequently filled (such as with copper). It is then planarized using a chemical mechanical polish. The dual damascene process combines pattern and vertical connections into one fabrication process.

Numerical aperture – A value that characterises the solid angle a sensor or light source is exposed to of an object, in this case, an ion. This is a dimensionless quantity.

Die – A cut-out piece of a larger, fabricated wafer. For integrated circuits, a die is typically packaged in an epoxy afterwards, making them incompatible with ultra-high vacuum environments.

Collection efficiency – The percentage of collected photons that have been emitted by an object.

Node – An established fabrication method which uses fixed fabrication methods to create structures. These can be characterised by feature size, materials, voltages and many other aspects.

Ball-grid array – A method that uses bumped balls to attach a die to a circuit.